\documentclass[a4paper]{jpconf}
\usepackage{graphicx,amsmath,hyperref}

\def\p{{\bf p}}
\def\q{{\bf q}}
\def\x{{\bf x}}
\def\P{{\bf P}}
\def\mL{\mathcal{L}}
\def\pbp{\bar\psi\psi}

\newcommand{\Eq}[1]{{Eq.~({\ref{#1}})}}
\newcommand{\Fig}[1]{{Fig.~{\ref{#1}}}}
\newcommand{\beq}{\begin{equation}}
\newcommand{\eeq}{\end{equation}}

\begin{document}

\title{Spatially modulated chiral condensates}

\author{Stefano Carignano}

\address{INFN, Laboratori Nazionali del Gran Sasso, Via G. Acitelli, 22, I-67100 Assergi (AQ), Italy}

\ead{carignano@lngs.infn.it}

\begin{abstract}
I discuss some properties of spatially dependent chiral condensates, focusing on one-dimensional modulations. 
After briefly introducing a generic formalism for studying inhomogeneous solutions in dense quark matter, 
I consider a plane-wave and a sinusoidal ansatz, and discuss the relationship between their order parameters and free energies, 
as well as a comparison of their dispersion relations. 
I then introduce a real kink crystal, and discuss its relationship with the sinusoidal ansatz by performing a Fourier analysis of this kind of solution.
\end{abstract}

\section{Introduction}

\begin{figure}[h!]
\begin{center}
\includegraphics[angle=270,width=.5\textwidth]{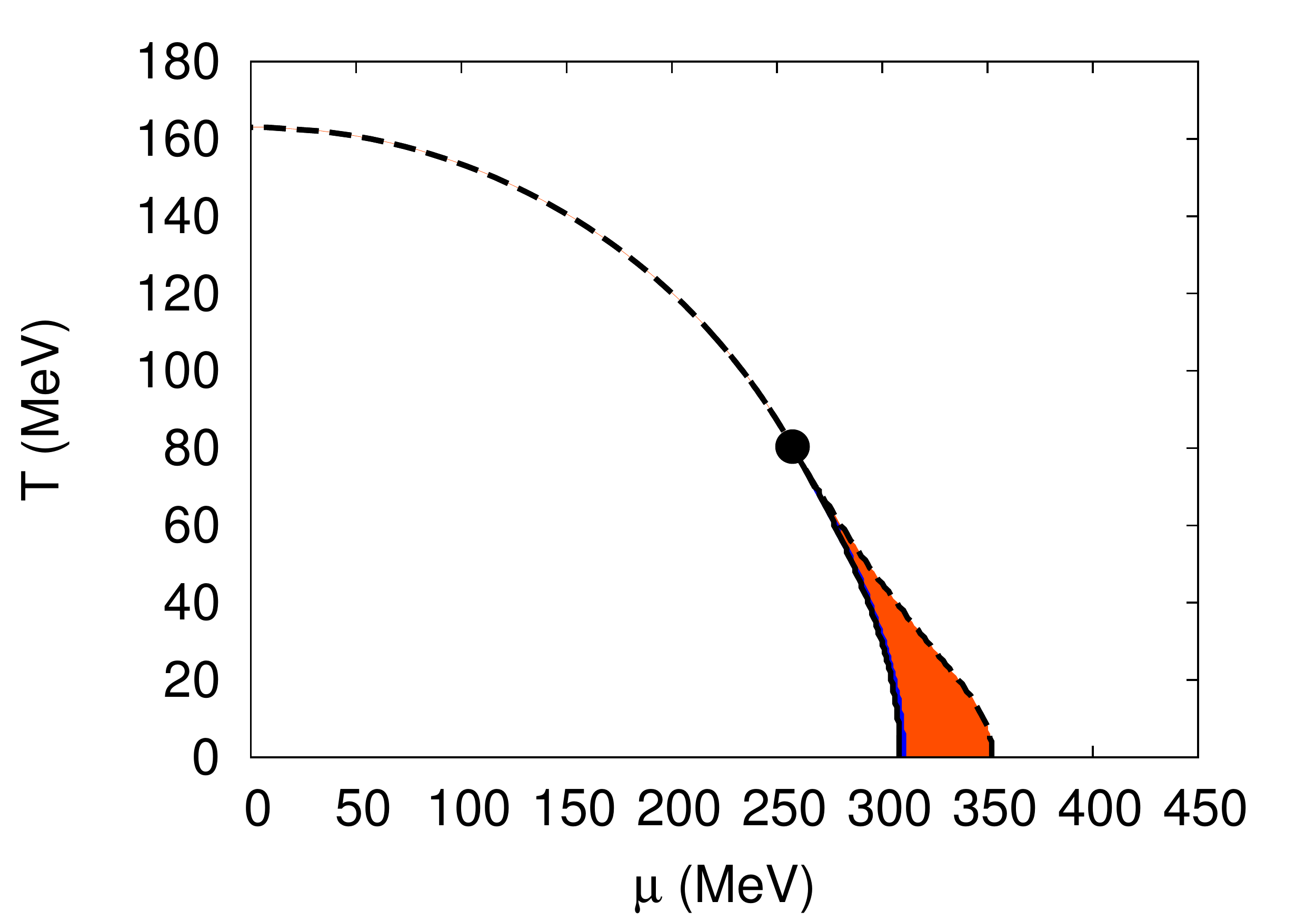}
\end{center}
\caption{\label{fig:pd}Phase diagram obtained within NJL model calculations when allowing for inhomogeneous chiral condensates. The shaded area 
denotes a region where spatially modulated chiral condensates are favored over the homogeneous chirally broken and restored solutions.  The actual 
size of the inhomogeneous window depends on the ansatz considered for the spatial dependence of the chiral condensate, as well as on model parameters.}
\end{figure}

While the properties of quantum chromodynamics (QCD) at finite density are still poorly known, 
its phase structure is expected to be extremely rich. 
Matter at low temperatures and asymptotically large baryonic chemical potentials should behave
as a color-superconductor, but other pairing mechanisms might become competitive as the density of the system decreases.
In particular, in the past few years some growing consensus has been building around the idea that
at densities around a few times nuclear matter saturation density and low temperatures,
spatially inhomogeneous chiral condensates might form, giving rise to crystalline phases. Explicit model calculations 
indeed corroborate this hypothesis by suggesting that an inhomogeneous ``island'' appears in this region, 
where the favored structure for the chiral condensate is a spatially modulated one (for a recent review, see \cite{Buballa:2014tba}). 
A typical phase diagram obtained within this kind of calculations is shown in \Fig{fig:pd}.

The phenomenon of inhomogeneous chiral symmetry breaking at finite densities stems from pairing of
quarks and holes with equal momenta on patches of the Fermi surface. The formed 
pairs thus carry a nonvanishing total momentum 
$|\P_{pair}| \sim 2\mu$, ($\mu$ being the quark chemical potential) and the corresponding condensate is spatially non-uniform.
 Superposition of paired quarks on different patches of the Fermi surface 
can then in principle give rise to more involved crystalline structures  \cite{Kojo:2011}.

Recent calculations performed within QCD-inspired effective models have shown that the favored shape for the 
chiral condensate in the inhomogeneous island is that of a modulation which varies in only one spatial direction \cite{Carignano:2012sx,Abuki:2011pf}. 
In this contribution, I will discuss a few properties of some of these one-dimensional solutions within an effective model of QCD.

\section{NJL model studies of inhomogeneous chiral condensates}
The theoretical framework employed in this work for the characterization of inhomogeneous chiral condensates in dense quark matter is the
 Nambu--Jona-Lasinio (NJL) model,
a popular tool often employed in the study of low-energy properties of strong-interaction matter
as well as its phase diagram 
at finite temperature and density.
In its simplest form, the NJL-model Lagrangian reads \cite{NJL2}
\beq
\mL_\text{NJL} = \bar\psi (i\gamma^\mu\partial_\mu -  m )\psi 
+ G_S \left( (\pbp)^2 + (\bar\psi i \gamma^5 \tau^a \psi)^2\right) \,,
\label{eq:LNJL}
\eeq
where $\psi$ represents a quark field with $N_f=2$ flavor and $N_c=3$ 
color degrees of freedom,
$\gamma^\mu$ and $\gamma^5$ are Dirac matrices, 
and $\tau^a$ $(a = 1,2,3)$ are isospin Pauli matrices.
The  $SU(2)_L\times SU(2)_R$ chiral symmetry of $\mL_\text{NJL}$  is explicitly broken by the bare quark mass $m$, 
as well as spontaneously by non-vanishing condensates of the 
form $\phi_\sigma \equiv \langle\pbp\rangle$ or $ \phi_{\pi}^a \equiv \langle\bar\psi i \gamma^5 \tau^a \psi\rangle$.

In the following I will consider the chiral limit, $m=0$, and work within the standard mean-field approximation where fluctuations of the order parameters are neglected.
 This allows to obtain the model thermodynamic
potential per volume $V$ for a given temperature and chemical potential, as a function of the scalar and pseudoscalar mean-fields:
\beq
\label{eq:Omega}
\Omega(T,\mu; \phi_\sigma , \phi_\pi^a) 
= -\frac{T}{V} \mathbf{Tr}\, \mathrm{Log} \left(\frac{S^{-1}}{T}\right)
+
\frac{T}{V}\int_{V_4} d^4x_E\; G_S (\phi_\sigma^2 + \phi_{\pi,a}^2)
\eeq
where the integral is performed in
Euclidean space-time, $x_E = (\tau, \x)$ with the imaginary time $\tau
= it$, and extends over the four-volume $V_4 = [0,\frac{1}{T}] \times V$.
The functional trace of the inverse quark propagator $S^{-1}$ runs over $V_4$ and internal (color, flavor, and Dirac) degrees 
of freedom. 

In order to investigate inhomogeneous chiral-symmetry breaking, within this approach the mean-fields are allowed to be spatially dependent, but are still assumed to be static 
so that the temporal part of the functional trace can be evaluated as a standard sum over fermionic Matsubara frequencies. 
%$\phi_\sigma = {\phi_\sigma}(\x)$ and ${\phi_{\pi}^a} = {\phi_{\pi}^a}(\x)$.

From a technical point of view, the inversion of the quark propagator requires the diagonalization of the effective Hamiltonian 
\beq
\label{eq:Hq}
 {\cal{H}} = \gamma^0 \left[ -i\gamma^i\partial_i - 2G_S (\phi_\sigma(\x) + i\gamma^5\tau_a \phi_{\pi}^a(\x)) \right] \,.
 \eeq
If charged pion condensation is neglected ($\phi_{\pi}^1  = \phi_{\pi}^2 = 0 \,, \phi_{\pi}^3 \equiv \phi_\pi $), after expliciting the Dirac structure it is possible to 
 factorize the Hamiltonian into a direct product of two isospectral blocks in flavor space \cite{Nickel:2009wj},
 \beq
 \label{eq:Hp}
 {\cal H}  = H_+(M) \otimes H_+(M^*) \,, \quad  \text{with} \qquad
 H_+(M)  =
  \left( \begin{array}{cc}   i \sigma^i \partial_i & M(\x) \\ M^*(\x) &  -i \sigma^i \partial_i \end{array} \right)  \,,
\eeq
 where $\sigma^i$ are Pauli matrices and one customarily defines a constituent mass function 
 \beq
 M(\x) = - 2G_S (\phi_\sigma(\x) + i  \phi_\pi(\x)) \,.
 \eeq

In presence of spatially modulated condensates, the diagonalization of ${\cal{H}}$ becomes extremely involved, as quark momenta are no longer conserved
and the Hamiltonian is not diagonal in momentum space. 
In practice, it is then useful to restrict the analysis to periodic modulations of the chiral condensate, an assumption which allows to perform a 
 Fourier decomposition of the mass function:
\beq
\label{eq:Mxq}
       M(\x) = \sum_{\q_k} \,\Delta_{k} e^{i\q_k \cdot \x}\,,
\eeq
with momenta $\q_k$ forming 
 a reciprocal lattice (RL) in momentum space.
Due to the crystal symmetries, only quark momenta differing by elements of the RL are paired in ${\cal{H}}$, 
so that one can effectively diagonalize a Hamiltonian with a discrete momentum structure.
The components of one of its momentum blocks are given by
\beq
[H_+]_{\p_m,\p_n} =
 \left( 
\begin{array}{cc}
 -\vec\sigma\cdot\p_m\,\delta_{\p_m,\p_n} &  
 \sum\limits_{\q_k} \Delta_{k} \delta_{\p_m,\p_n+\q_k} 
 \\
 \sum\limits_{\q_k} \Delta^*_{k} \delta_{\p_m,\p_n-\q_k} &  
 \vec\sigma\cdot\p_m\,\delta_{\p_m,\p_n} 
\end{array} 
\right) \,,
\label{eq:Hmn}
\eeq
where it can be clearly seen that the inhomogeneous chiral condensate couples different quark momenta. %, leading to a non-diagonal structure.

Any periodic modulation of arbitrary spatial dimension can in principle be implemented within this approach by fixing an underlying lattice structure via the definition of the $\q_k$ and considering
 an appropriate number of Fourier coefficients.
The favored solutions can then be obtained by minimizing the model thermodynamic potential with respect to the variational parameters $\lbrace \Delta_k, \q_k \rbrace$.
In practice, however, the numerical diagonalization of the quark Hamiltonian in momentum space becomes extremely demanding from a numerical point of view
 as the mass ansatz becomes more involved.  A dramatic simplification of the problem comes if one considers solutions for which the chiral condensate is  modulated 
 only along one spatial direction (which one can take without loss of generality along the $z$ axis),
 while remaining constant along the others. In particular, it was shown in \cite{Nickel:2009wj} that it is usually possible to solve the eigenvalue problem  by diagonalizing a dimensionally-reduced Hamiltonian with $p_x = p_y = 0$
  and subsequently boost the obtained energies along the transverse directions.
  
  A systematic NJL model analysis on the favored type of modulation including higher-dimensional ans\"atze has been performed in \cite{Carignano:2012sx} at zero temperature, 
  and in \cite{Abuki:2011pf} in proximity of the chiral critical point within a Ginzburg-Landau expansion study. Both agree on finding that the thermodynamically favored solution is a one-dimensional one, in particular a so-called real kink crystal (RKC) which can be expressed in terms of Jacobi elliptic functions. This is in contrast with the situation for the 1+1-dimensional NJL model (NJL$_2$), where the ground state is given by a plane wave modulation of the chiral condensate often referred to as ``chiral spiral'' \cite{Schon:2000qy}. On the other hand, RKC solutions 
  are found to be favored in the Gross-Neveu model \cite{Schnetz:2004}, which has only a discrete chiral symmetry (see also \cite{Kojo:2014fxa} for a comparison of the pairing mechanisms in the two models). 
  From a technical point of view, it is the particular form of the Hamiltonian in the NJL model in 3+1 dimensions (\Eq{eq:Hp}) containing two blocks differing only by the substitution $M \leftrightarrow M^*$ which makes these real solutions the favored ones  \cite{Nickel:2009wj}.
  
\section{Comparison of one-dimensional modulations} 

As previously mentioned, in order to perform a model analysis of inhomogeneous condensation, a given ansatz for the spatial dependence 
of the mean-fields must be chosen. 

Among all possible one-dimensional solutions, the simplest is a so-called ``(dual-)chiral density wave'' (CDW) \cite{NT:2004},
a plane wave ansatz which can be seen as analogous of the Fulde-Ferrell solutions 
introduced in the context of superconductivity~and color superconductivity:

\beq
\label{eq:Mcdw}
M_{CDW}(\x) = \Delta e^{i q z} \,, 
\eeq
with two variational parameters $\Delta$ and $q$, characterizing the amplitude and the wave number
of the modulation.  This can be seen as a one-dimensional truncated version of the general ansatz of \Eq{eq:Mxq} where only the first Fourier component $\Delta_1$ is allowed to be nonzero. 

A very important feature of the CDW is 
that for this ansatz the explicit spatial dependence of the condensates in the quark Hamiltonian (\Eq{eq:Hq}) can be removed by means
of a local chiral rotation of the quark fields \cite{Dautry:1979,Kutschera:1989yz} (see also \cite{Buballa:2014tba}),
thus allowing to evaluate the eigenvalue spectrum analytically.
The resulting eigenvalues of the dimensionally reduced Hamiltonian are the positive and negative square roots of
\beq
\label{eq:CDWdispers2}
        E^2_\pm(p_z) = p_z^2 + \Delta^2 + \frac{q^2}{4} \pm\sqrt{\Delta^2 q^2 + (q p_z)^2} = \left(\sqrt{p_z^2 + \Delta^2}  \pm \frac{q}{2}
       \right)^2  \,.
\eeq

An alternative ansatz often considered in literature is a simple real sinusoidal modulation of the mass function, for which the pseudoscalar part $\phi_\pi$ is set to zero: 
\beq
M_{COS}(\x) = \Delta \cos(q z)  \,.
\eeq
From a technical point of view, this ansatz can be seen as a superposition of two plane waves propagating in opposite directions, and as such its spatial dependence
cannot be rotated away using the technique employed for a CDW. In fact, no analytical expression for the eigenvalue spectrum in presence of this modulation is known,
so that a numerical diagonalization of the model Hamiltonian in momentum space must be performed, using, for example, the techniques developed in \cite{NB:2009,Carignano:2012sx}.

\begin{figure}
\begin{center}
\includegraphics[width=.46\textwidth]{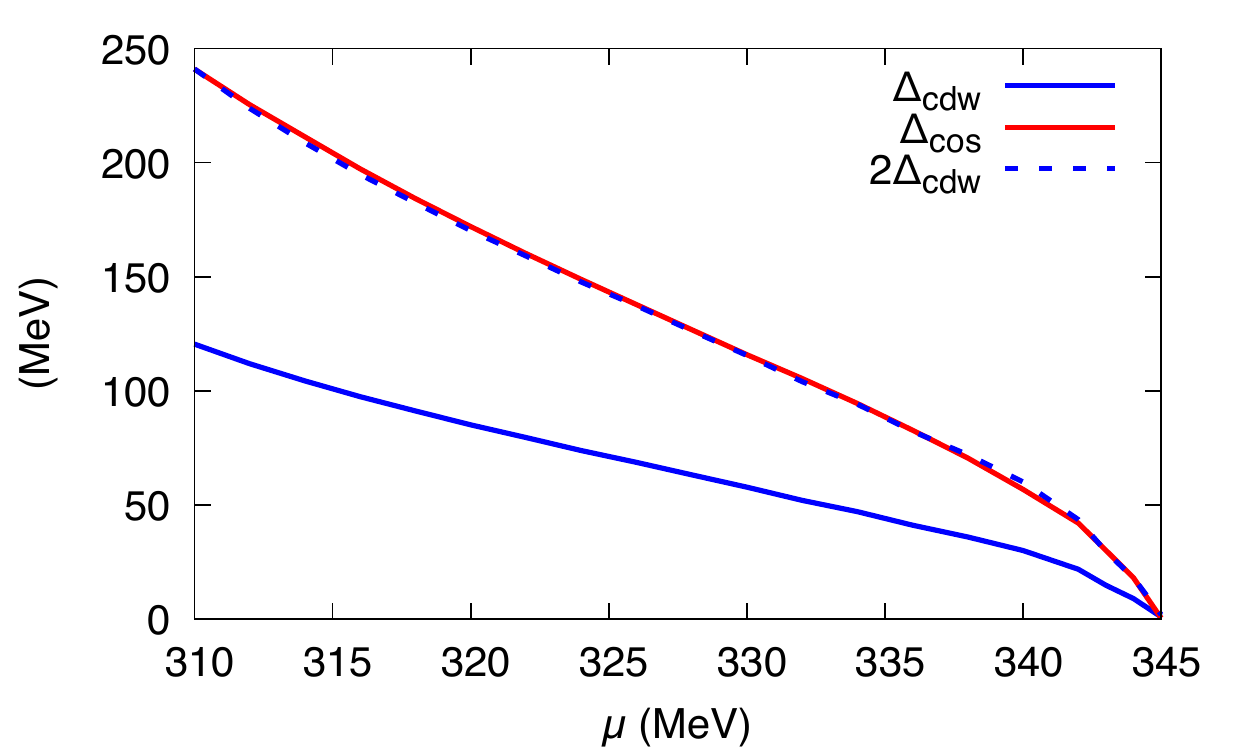}
\includegraphics[width=.46\textwidth]{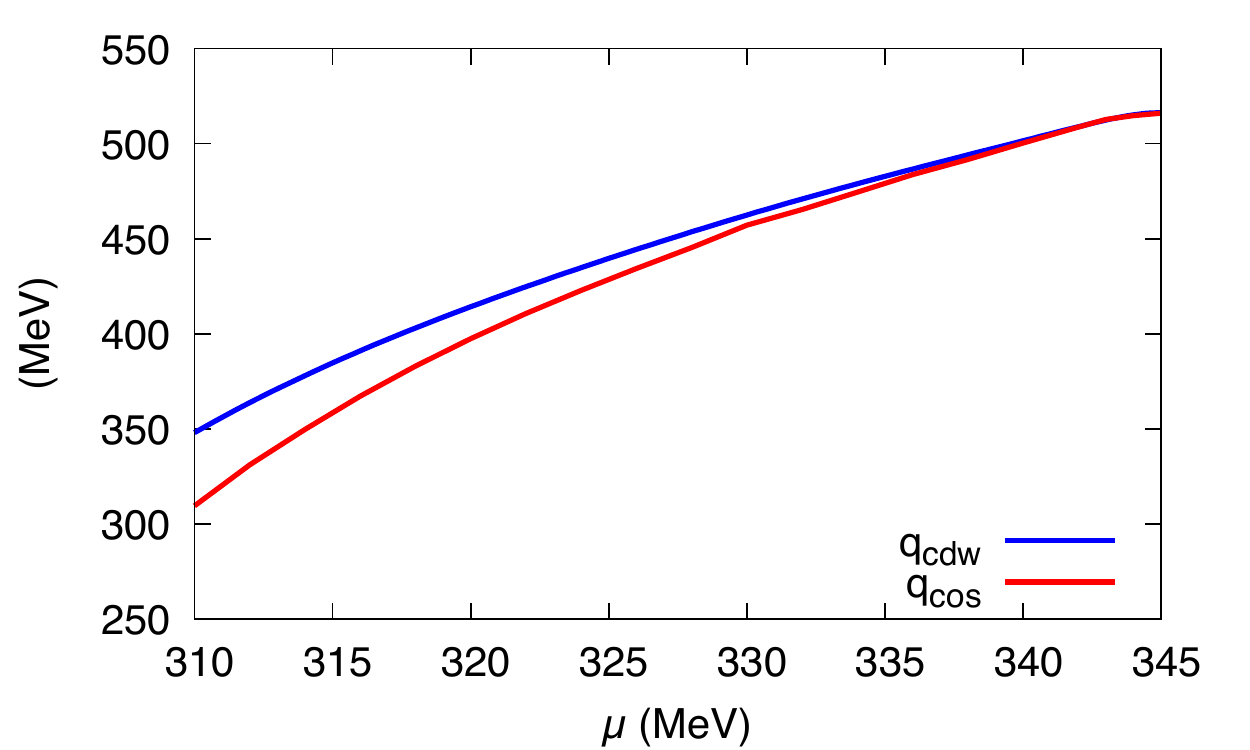}
\end{center}
\caption{\label{fig:ops}Left: comparison of amplitudes at $T=0$ for CDW and COS modulation. The solid blue line denotes the thermodynamically favored $\Delta$ for the CDW modulation, the dashed blue line
 is twice that value, the solid red line is the favored $\Delta$ for the COS modulation.   Right: comparison of wave numbers. The blue line is the favored  $q$ for the CDW, the red one is the $q$ for the COS. }
\end{figure}

\begin{figure}
\begin{center}
\includegraphics[width=.5\textwidth]{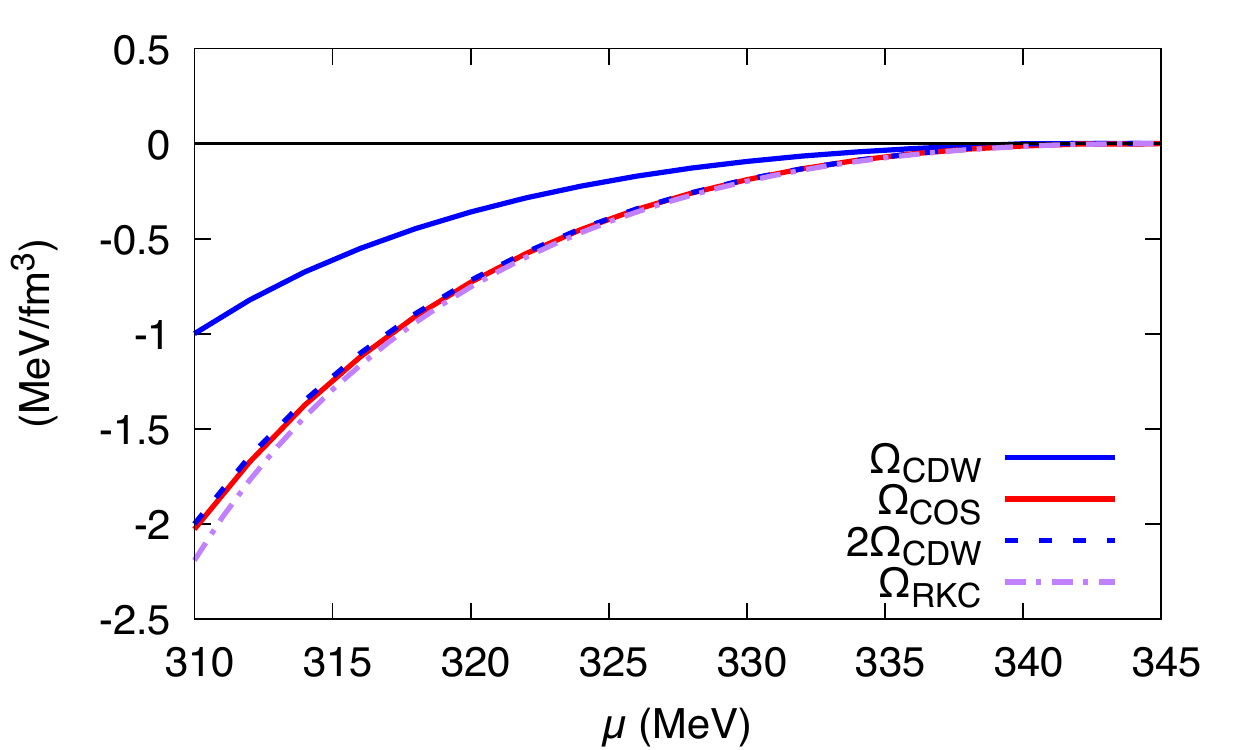}
\end{center}
\caption{\label{fig:omega}Free energy associated with different modulations of the chiral condensate in the region where inhomogeneous phases are thermodynamically favored, as a function of 
quark chemical potential ($T=0$).
 Solid blue line: CDW, Red line: COS. 
The dashed blue line, corresponding to twice the CDW free energy, is basically overlapping the COS one throughout the whole inhomogeneous window. The purple dash-dotted line denotes 
the RKC free energy, which is slightly lower than the COS one for lower $\mu$ but rapidly becomes degenerate with it. } 
\end{figure}

After minimization of the thermodynamic potential, one can observe a very close relationship between the order parameters and the free energies of the CDW and COS modulations. 
Indeed, as can be seen in \Fig{fig:ops}, throughout almost the entire inhomogeneous window one finds that $\Delta_{COS} = 2\Delta_{CDW}$, while $q_{COS}$ and $q_{CDW}$ are similar but not equal and start
overlapping only close to the chiral restoration phase transition, a behavior already predicted within a GL analysis in \cite{Nickel:2009wj}.  Another interesting result is that the COS solution is always favored 
by a factor of 2 in free energy compared to the CDW one, as shown in \Fig{fig:omega}.

\begin{figure}
\begin{center}
\includegraphics[width=.44\textwidth]{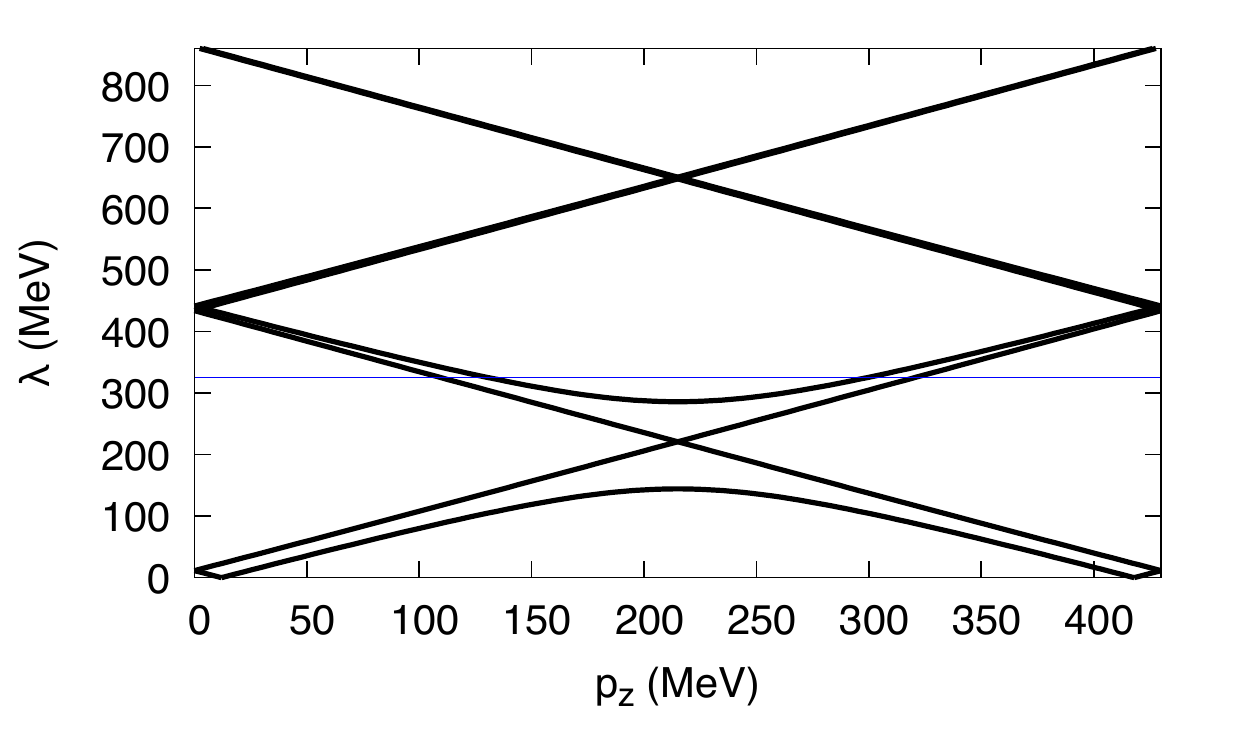}
\includegraphics[width=.44\textwidth]{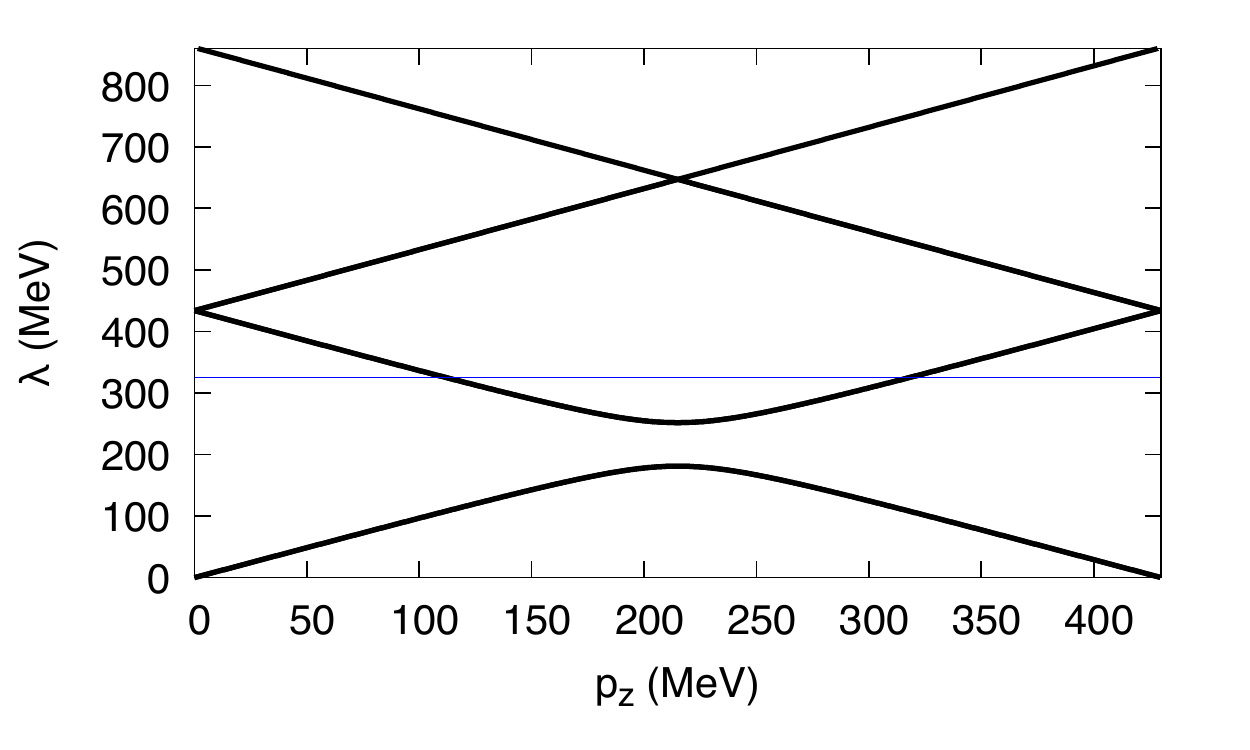}
\end{center}
\caption{\label{fig:disprel} Dispersion relations for the CDW (left) and COS (right) modulations. Both have been evaluated for values of the order parameters corresponding to the
respective minima at $T=0,\mu=325$ MeV (recall at the minimum $q_{COS} \simeq q_{CDW}$ and $\Delta_{COS} \simeq 2\Delta_{CDW}$). The chemical potential is plotted as reference as an horizontal blue line.}
\end{figure}

In order to understand why the real COS solution is favored over the complex CDW one, it might be also of interest to compare the dispersion relations for the two modulations. 
The eigenvalues of the dimensionally-reduced Hamiltonians for the two are plotted in \Fig{fig:disprel}. There it can be seen how the CDW spectrum still exhibits branches 
which are analogous to those of chirally restored quark matter, whereas for the COS modulation these also acquire a gap. This additional pairing present for the COS modulation allows for a larger energy gain than in the CDW case, since twice as many branches become deformed and allow for the system to further lower its energy.  

The final part of my discussion is devoted to the thermodynamically favored modulation in the NJL model, which is known to be given by a real-kink crystal. This particular kind of ansatz can be expressed as~\cite{Schnetz:2004} (see also \cite{Buballa:2014tba}) 
\beq
\label{eq:Mzsolitonsalt}
M(z) = \Delta \sqrt{\nu}\, \text{sn} (\Delta z \vert \nu) \,,
\eeq
where sn is a Jacobi elliptic function and the parameter $\nu$, called elliptic modulus and ranging from 0 to 1, describes the shape and periodicity of the modulation.  

By inspecting the behavior of the mass function as chemical potential increases, one can see that at the onset of the inhomogeneous phase this solution has the limit $M(z) \to \Delta \tanh(\Delta z)$, which can be seen as a single soliton interpolating between the two homogeneous chirally broken solutions.  As $\mu$ increases, solitons start stacking up against each other and the order parameter progressively assumes a sinusoidal shape \cite{CNB:2010,Buballa:2012vm}.  This kind of solution and its behavior can be encompassed within the framework introduced in the previous section by evaluating the Fourier coefficients associated with it, as function of chemical potential.  The resulting values are presented in Table \ref{tab:highharm1d}.  There one can see, as expected, that as the chemical potential increases the higher Fourier coefficients rapidly drop to zero, and the RKC solution assumes the form of a simple cosine.

\begin{table}[h]
  \caption{
 \label{tab:highharm1d}
 Values of the higher order nonzero harmonics
    for a RKC modulation
    at different values of chemical potential
    close to the onset of the inhomogeneous phase ($T=0$).
    Results are normalized to the value of the first harmonic ($n=1$) in order to show their
  relative weights. Being a real modulation, one also finds $\Delta_{-n} = \Delta_n$. Even coefficients are found to be zero.}
  
  \begin{center}
  \begin{tabular}{c c c c c c}
    \br
    $\mu$  & $\Delta_3/\Delta_1$ & $\Delta_5/\Delta_1$ &  $\Delta_7/\Delta_1$  &  $\Delta_9/\Delta_1$     \\
    \mr
    308.0 & -0.12936 & 0.01959 & -0.002978 & 0.000452  \\
    310.0 &  -0.04672 & 0.00229 & -0.000112 & 0.000005  \\
    315.0 & -0.02057 & 0.00043 & -0.000009 & 0.000000  \\
    320.0 &  -0.01174 & 0.00013 & -0.000001 & 0.000000  \\
    \br
  \end{tabular}
  \end{center}

\end{table}

\section{Conclusions}

In this contribution I discussed some properties of one-dimensional modulations of the chiral condensate in dense quark matter 
within NJL model calculations. In particular, I focused on two  possible one-dimensional solutions, namely a chiral density wave and a single cosine, 
which, in the framework of a Fourier expansion of the order parameter, can be seen as prototypes for complex and real ans\"atze, respectively.   
  I showed 
  that the thermodynamically favored values for the wave numbers of the chiral density wave and that of the cosine are very similar throughout the whole 
inhomogeneous window, while the amplitude of the cosine is twice that of a plane wave. 
Furthermore, by inspecting the quasiparticle dispersion relations I argued why a real sinusoidal modulation is favored over a complex plane wave one. In particular, 
the fact that for the COS modulation twice as many branches as for the CDW become gapped could explain the fact that the energy gain for the former 
is twice the one of the latter.

Finally, I considered the solution which is found to be the most favored throughout the whole inhomogeneous window, a real kink crystal, 
and investigated its relationship with a simpler real sinusoidal modulation by performing a Fourier decomposition of this type of ansatz. The RKC 
effectively reduces to a simple cosine rapidly after the onset of the inhomogeneous phase.  This in turn implies that for most intents and purposes the RKC solution can be 
reasonably approximated by considering a single cosine, with the exception of the immediate vicinity of the phase transition associated with the onset of inhomogeneous condensation.
In particular, it is important to remember that the nature of such phase transition depends strongly on the type of modulation considered. Among those discussed here,
only the RKC can smoothly reduce  in the thermodynamic limit to an homogeneous chirally broken solution. Indeed, in order to reproduce an homogeneous solution which has an infinite period, an arbitrarily large 
set of Fourier modes needs to be considered.  As such, only the RKC solution can realize a second-order transition to the homogeneous chirally broken phase, whereas for the other simpler 
 ans\"atze a first-order transition must occur. 

\section{Acknowledgments}
I would like to thank M.~Buballa for helpful discussions.

\section*{References}

\providecommand{\newblock}{}

\end{document}